\documentclass[useAMS,usenatbib]{mn2e}

\title[Variable stars in NGC~6397]
{Photometric study of the variable star population in the globular
cluster NGC~6397}
\author[J. Kaluzny et. al.]
{J.~Kaluzny$^{1}$, I.~B.~Thompson$^{2}$, W.~Krzeminski$^{3}$, A.
Schwarzenberg-Czerny$^{1,4}$\\
  $^1$Nicolaus Copernicus Astronomical Center,
     ul. Bartycka 18, 00-716 Warsaw, Poland
     (jka,alex@camk.edu.pl)\\
  $^2$Carnegie Institution of Washington, 813 Santa Barbara Street,
        Pasadena, CA 91101, USA
        (ian@ociw.edu)\\
  $^3$Las Campanas Observatory, Casilla 601, La Serena, Chile,
(wojtek@lco.cl)\\
  $^4$Adam Mickiewicz University Observatory, ul. Sloneczna 36,
  61-287 Poznan, Poland
\\
\\
}

\date{Accepted .......
      Received .......;
      in original form ........}

\pagerange{\pageref{firstpage}--\pageref{lastpage}} \pubyear{2004}

\begin{document}

\maketitle

\label{firstpage}

\begin{abstract}
We present the results of a photometric survey for variable stars
in the central region of the nearby globular cluster NGC~6397.
Time series photometry was obtained for 30 variable objects. The
sample includes 12 new objects,  of which 6 show periodic light
curves and 2 are eclipsing binaries of unknown period. Six
variables possess certain and three possess likely X-ray
counterparts detected with the $Chandra$ observatory. Among them
four are cataclysmic variables and one is a foreground eclipsing
binary. The cataclysmic variable CV2 exhibited a likely dwarf nova
type outburst in May 2003. The cataclysmic variable CV3 was
observed at $18.5<V<20.0$ during 5 observing runs,  but  went into
a low state in May 2003 when it reached $V>22$. We have found that the
light curve of the optical companion to the millisecond pulsar PSR
J1740-5340 exhibits noticeable changes of its amplitude on a time
scale of a few  months. A shallow eclipse with $\Delta V=0.03$~mag
was detected in one of the cluster turnoff stars suggesting the
presence of a large planet or brown dwarf in orbit.
\end{abstract}
\

\section{Introduction} \label{s1}
The central region of the post-core-collapse cluster NGC 6397 is
known to contain several objects likely to have been  created as a
result of stellar interactions in its core region. Identification
of three candidate cataclysmic variables (CVs) based on  $HST$
data (Cool et al. 1995; Grindlay et al. 1995) was followed by
detection of 25 X-ray sources with the $Chandra$ observatory
(Grindlay et al. 2001). The $Chandra$ sources include 8 candidate
or confirmed CVs, a quiescent low mass X-ray binary and a
millisecond pulsar PSR~J$1740-5340$. The sample of variable stars
known in the cluster field contains several pulsating SX~Phe stars
and contact binaries (Kaluzny 1997; Kaluzny \& Thompson 2003,
hereafter KT03).

In this paper, we report the results of a new photometric survey aimed
mainly at the identification of detached eclipsing binaries in the
central part of NGC~6397.
Such objects can potentially serve as distance and age indicators for the
cluster itself (Paczy\'nski 1997). The reported survey is a part of the 
long-term CASE project, which has been conducted for several years
at Las Campanas Observatory (Kaluzny et al. 2005).
Secondary goals of the reported observations  included detection of
optical variability and a search for possible outbursts of CVs candidates
and examination of the stability of the light curve of an optical
counterpart to the millisecond pulsar (Ferraro et al. 2001).

\section{Observations}\label{s2}
All images were taken at Las Campanas Observatory with the 2.5-m
duPont telescope.  A field of $8.65\times 4.33$ arcmin$^{2}$ was
observed with the TEK No. 5 CCD camera at a scale of 0.259
arcsec/pixel. The cluster core was positioned 32 arcsec West of
the field center. Observations were made on 14 nights during 4
observing runs conducted from May 2003 to June 2004. Images were
taken in two bands with exposure times 10-20~s for the $V$ and
20-40~s for the $B$ filters. During the 2003 season observations
were limited to the $V$ band only. In total, 1951 frames in $V$ and
288 frames in $B$ were collected and the field was monitored for
55 hours.

The raw data were pre-processed with the
IRAF-CCDPROC package.\footnote{
IRAF is distributed by the National
Optical Astronomy Observatories, which are operated by the Association
of Universities for Research in Astronomy, Inc., under cooperative
agreement with the NSF.}
Groups of 3-5 consecutive frames were co-added, and the total number
of stacked images used in
the present study was 375 and 85, for $V$ and $B$ bands respectively.
The resultant time resolution of our photometry based on combined images is
about 5 minutes. The median value of seeing was 1.01 and 1.02 arcsec for
the $V$ and $B$ band, respectively.

\section{Photometry and identification of variables}\label{s3}
To search for variable objects we used the ISIS-2.1 image subtraction
package (Alard \& Lupton 1998; Alard 2000). Our procedure
followed closely that described in some detail by Mochejska et al.
(2002). Instrumental magnitude zero points for the ISIS differential
light curves were measured from the template images, using the DAOPHOT/ALLSTAR
package (Stetson 1987), and aperture corrections were measured with the
DAOGROW program (Stetson 1990). Instrumental magnitudes were transformed
to the standard $BV$ system using a large assembly of local standards
established during our earlier study of the same cluster (Kaluzny,
Ruci\'nski \& Thompson 2003). The monitored field includes 6 stars from the
Stetson (2000) photometric catalog; standard $V$ magnitudes are
available for all of them. We find mean difference
$\Delta V=0.036\pm 0.006$, in the sense that our magnitudes are brighter.
$B-V$ colors are available  for only two Stetson stars located in
our field. We obtained $\Delta (B-V)=-0.031$ and $\Delta (B-V)=-0.032$
for Stetson objects \#86 and \#88, respectively (our colors are redder).
Consequently we have applied corrections of $\Delta V=0.036$ and
$\Delta (B-V)=-0.031$ to our photometry. All photometric quantities
listed below include these corrections.

Two methods were used to detect variable objects. The first method
relies on examination of the light curves for all stellar objects
detectable  with the profile photometry on the reference image for
the $V$ band. The PHOT procedure in the ISIS package was used to
extract differential light curves at the position corresponding to
each stars detected with DAOPHOT/ALLSTAR. The light curves were
then transformed to magnitudes and checked for variability using
the TATRY program written by ASC. This program is suitable for
detection of various types of variable stars and  in
particular uses the AoV algorithm (Schwarzenberg-Czerny 1996) to search
for the presence of any periodic signal in the analyzed light curves. The
second method is based on examination of residual images produced
by the ISIS package. In particular it allows  searching for unresolved
variables which were not detected with the profile photometry.

A total of 30 variables were identified. Their equatorial coordinates
are listed in Table 1. The astrometric frame solution was derived
using the coordinates of 547 stars extracted from the GSC-2 catalog
with the help of the Skycat-2.7.3 interface. Names listed in column (1)
of Table 1 follow Clement et al. (2001) for objects V4-24.
Variables V25-36 are new identifications. Column (2) gives alternate
designations used in Grindlay et al. (2001). Columns (5-9)
list some basic photometric properties of variables deduced from our
photometry. Most of the objects can be classified as periodic variables.
For objects V4-24 the listed periods were derived from photometry
including observations from the 2002 season reported in KT03.

The finding charts
for variables V33=CV3 and V34=CV2 can be found in Cool et al. (1995).
Variable V31 can be unambiguously identified as an object located
at $(X,Y)=(475,241)$ on the HST/WFPC2-PC1 image u5dr0401r.
The finding charts for the remaining new variables are shown in Fig. 1.

Positions of the variables on the cluster color-magnitude diagram are shown
in Fig. 2.
\setcounter{figure}{1}
\begin{figure}
\begin{center}
\vspace{7.2cm}
\includegraphics{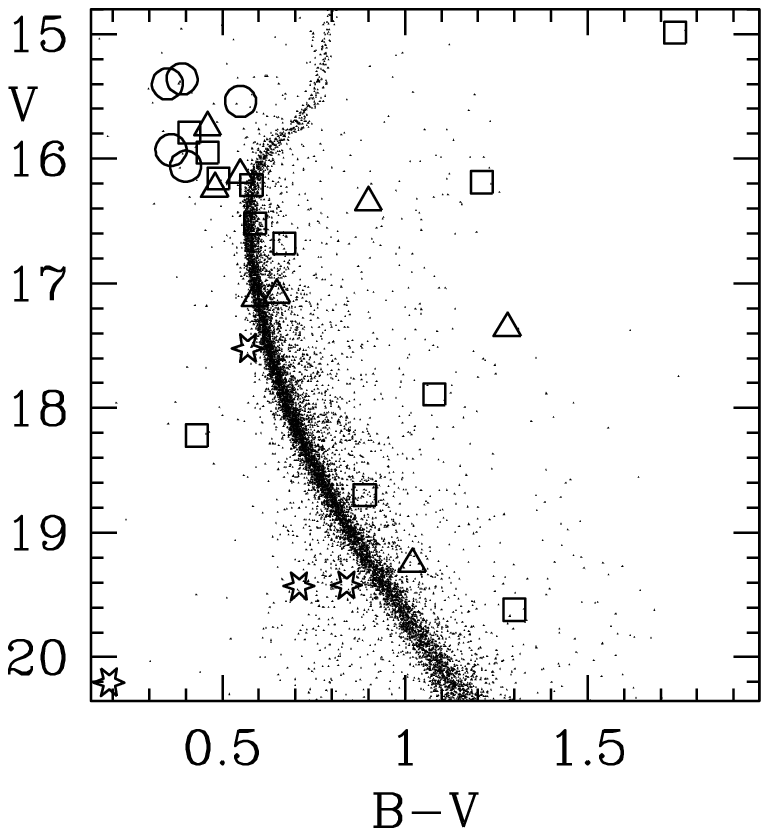}
\end{center}
\caption{\label{f2} Color-magnitude diagram for NGC~6397, with positions of
the variables marked. Triangles indicate eclipsing binaries,
circles pulsating stars, asterisks CVs, and squares the remaining variables.}
\end{figure}

\begin{table*}
 \centering
 \begin{minipage}{200mm}
  \caption{ Variable stars in the field of NGC~6397.\label{t1}}
{\small
  \begin{tabular}{llcclllll}
\hline
Name & ID& RA(2000)    & Dec(2000)             & Period & $V_{\rm max}$ & $V_{\rm min}$ & $<B-V>$  & Type/remarks\\
     &   & $~~h~~m~~s$ &  $~~~^\circ ~~~'~~~''$& days   &               &               &          &  \\
\hline
V4  & & 17 41 08.81 &  -53 42 34.0 &0.422740 &16.35 &17.08 &0.90 & eclipsing W UMa\\
V7  & & 17 40 43.89 &  -53 40 35.1 &0.269861 &17.09 &17.56 &0.65 & eclipsing W UMa\\
V8  & & 17 40 39.25 &  -53 38 46.8 & 0.271243 &16.24 &16.62 &0.48 & eclipsing W UMa\\
V10 & & 17 40 37.51 &  -53 40 36.1 & 0.030754  &15.93 &16.06 &0.36 & SX Phe\\
V11 & & 17 40 44.10 &  -53 40 40.4 & 0.038262 &15.40 &15.45 &0.35 & SX Phe\\
V12 &CV1, U23&17 40 41.57& -53 40 19.2&0.471199 &17.52 &18.01 &0.57 & CV \\
V13 &CV6, U10& 17 40 48.95& -53 39 48.5& 0.235196 &19.43 &19.80 &0.71 & CV\\
V14 & &17 40 46.46 &  -53 41 15.2 &0.33509  &19.25 &19.41 &1.02 & eclipsing   \\
V15 & &17 40 45.38 &  -53 40 24.8 &0.023824  & 15.40 &15.46 &0.35 & SX Phe  \\
V16 &MSP, U12&17 40 44.59& -53 40 41.4&1.35406  &16.68 &16.90 &0.67 & ellipsoidal \\
V17 & &17 40 43.78 &  -53 41 16.2 &1.061132   &16.21 &16.25 &0.58 & spotted?, ellipsoidal?  \\
V18 & &17 40 43.60 &  -53 40 27.6 & 0.786689 &15.75 &15.89 &0.49 & eclipsing \\
V19 & &17 40 42.81 &  -53 40 23.3 & 0.253755 &17.12 &17.22 &0.59 & eclipsing   \\
V20 & &17 40 41.66 &  -53 40 33.1 & 0.861160 &15.79 &15.87 &0.41 & $\gamma$ Dor?  \\
V21 & &17 40 41.55 &  -53 40 23.4 & 0.038961 &15.32 &15.62 &0.39 & SX Phe \\
V22 & &17 40 41.11 &  -53 40 41.9 & 0.344      &16.06 &16.24 &0.40 & RRd, P0=0.52 \\
V23 & &17 40 39.30 &  -53 40 46.5 & 0.037153 &15.51 &15.54 &0.55 & SX Phe \\
V24 & &17 40 39.07 &  -53 40 22.9 & 0.457176    &16.16 &16.18 &0.49 & $\gamma$ Dor? \\
V25 & &17 41 10.16 &  -53 39 30.5 & 1.2306?  &17.89 &18.06 &1.08 & periodic?\\
V26 &U42?& 17 40 43.05& -53 38 31.2& ?        &16.19 &16.28 &1.21 & irregular \\
V27 & &17 41 13.80 &  -53 41 14.1 &0.55614 &18.22 &18.27 &0.43 & ellipsoidal? \\
V28 & &17 41 02.70 &  -53 39 47.0 &25.997   &14.99 &15.27 &1.74 & pulsating? \\
V29 & &17 40 59.64 &  -53 40 38.6 &1.2434   &19.62 &19.81 &1.30 & ellipsoidal? \\
V30 &U5 &17 40 54.51& -53 40 44.4&?        &17.36 &17.92 &1.28 & eclipsing \\
V31 &U18?& 17 40 42.59& -53 40 27.1&1.30995  &15.95 &16.03 &0.46 & ellipsoidal? \\
V32 & &17 40 40.30 & -53 41 25.2 &?        &16.13 &16.16 &0.55 & eclipsing \\
V33 &U17, CV3& 17 40 42.62& -53 40 19.0&? &  18.3  & 22.5:  &  0.2:  & CV \\
V34 &U19, CV2& 17 40 42.29& -53 40 28.6&? &  16.2  & 22.0:  &  0.8:  & CV \\
V35 &U14? &17 40 43.31& -53 41 55.2&0.298438 &18.70 &18.87 &0.89 & ellipsoidal? \\
V36 & &17 40 44.10 &  -53 42 11.3 &0.54866   &16.52 &16.54 &0.59 & pulsating? \\
\hline
\end{tabular}}
\end{minipage}
\end{table*}
%\clearpage
% applied V38->V32   V37->V36
%
\setcounter{figure}{0}
\begin{figure*}
\begin{center}
\vspace{9.4cm}
\includegraphics{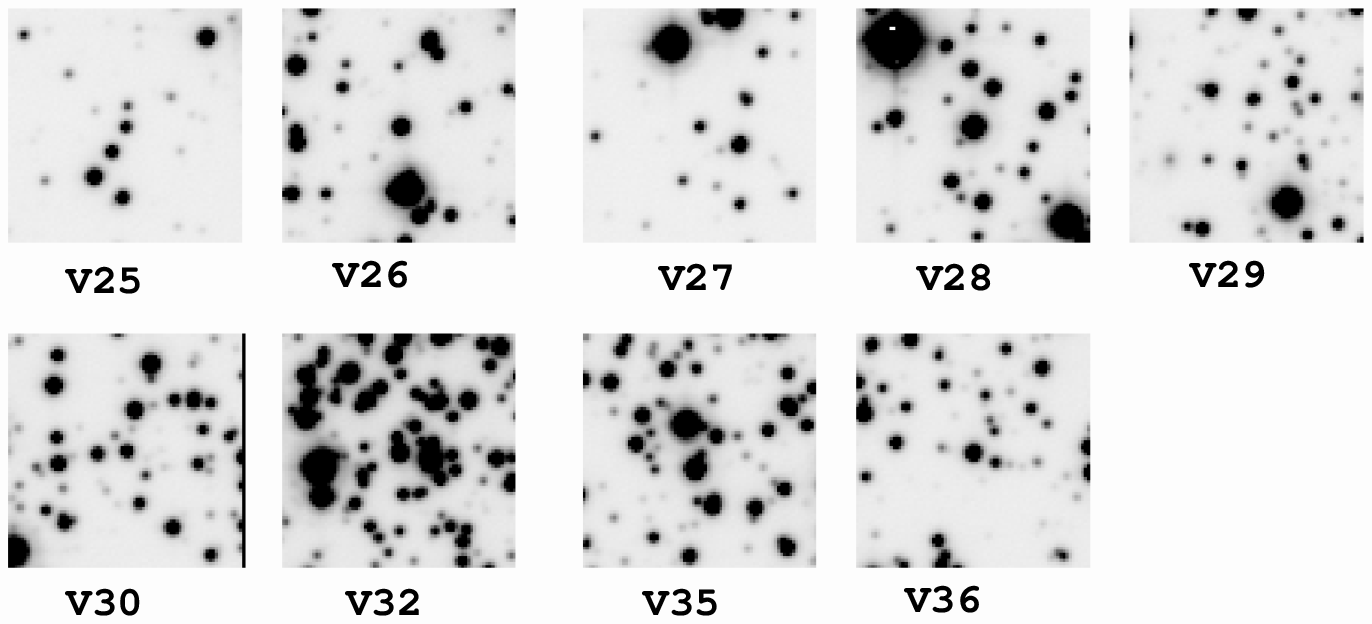}
\end{center}
\caption{\label{f1} Finder charts for some of newly identified variables.
Each chart is $20\arcsec$ on a side. North is up, and east to the left.
Variables are located right at the center of respective images.}
\end{figure*}
\subsection{$Chandra$ sources}
Based on  $Chandra$ deep-imaging observations, Grindlay et al. (2001)
detected 25 X-ray sources within $2\arcmin$ of the cluster center.
Five of the variables listed in Table 1 can be unambiguously identified as
optical counterparts to the $Chandra$ sources. These are cataclysmic
variables CV1-3 and CV6, and an optical companion to the millisecond
pulsar  PSR J1740-5340 (Ferraro et al. 2001). For these objects
we find mean difference of equatorial coordinates
$\Delta\alpha=0.09\pm 0.04\arcsec$ and $\Delta\delta=0.60\pm 0.06\arcsec$,
in the sense that $Chandra$ positions are to West and South of positions
derived from our frame solution. We have added the above listed offsets to
coordinates of other sources reported in Grindlay et al. (2001) and looked
for their possible counterparts on the list of detected variables.
Variable V30 can be considered as a very likely counterpart of the
$Chandra$ source U5. The difference of coordinates from the two surveys
amounts to 0.06~arcsec. For variables V26(U42), V31(U18) and V35(U14)
the differences amount to 0.35, 0.41 and 0.38~arcsec, respectively.
We note that source U18 was identified by Grindlay et al. (2001)
with  BY Dra candidate or possible millisecond pulsar based on analysis of
HST/WFPC2 images.

We also performed a search for possible variables by extracting,
with ISIS, light curves at locations corresponding to the $Chandra$ sources.
This procedure gave a positive result only for U21=CV4. A clearly variable
non-periodic object was detected at the predicted location of that source.
However, due to crowding problems it was impossible to derive
profile photometry for it and to transform its light curve from differential
count units  into magnitudes.
We may only note that Piotto et al. (2002) obtained for CV4  $V=20.68$
and $B=21.49$ based on  HST/WFPC2 imaging.
We did  not detect any variable sources at positions corresponding to
candidate CVs U22=CV5, U11=CV7, U13=CV8 and U28=CV (Grindlay et al. 2001).
Source U13 is located $0.8\arcsec$ from an SX~Phe variable V11, while
sources U22 and U28 are located very close to bright non-variable stars.

\section{Properties of variables}\label{s4}

\subsection{Cataclysmic variables}
Plots of $V$ magnitude against time for cataclysmic variables
CV1-3 and CV6 for the observing seasons 2002-2004 are shown in
Fig. 3. Stars V12=CV1 and V13=CV6 were identified as eclipsing CVs
in KT03. The light curves of both of them exhibit two distinct minima
per orbital period. The shape and mean level of the optical light
curve of CV6 was remarkably stable over three observing seasons.
Figure 4 shows its phased light curve for the 2004 season. It is
evident that the accretion rate in this system was roughly
constant over the last 3 years. Also, the light curve seems to be
dominated by ellipsoidal variations. The light curve of
variable V12=CV1 is less stable than that of CV6. The average $V$
magnitude of the variable increased by about 0.5 mag between the 2002
and 2004 seasons. The unstable character of the light curve of CV1
is demonstrated in Fig. 5, which shows phased light curves for 4
selected nights from the 2004 season. These light curves were
extracted from individual images and subsequently were smoothed
using a box-car filter with $N=3$. Effective time resolution is
about 80 seconds. Besides overall changes of the shape of the
eclipse, one may notice the presence of non-periodic oscillations
and/or flickering with $\Delta V$ reaching about 0.15 mag. Such
oscillations were not observed during the 2002 season. Our photometry
of CV1 and CV6 fails to reveal any periodicities other than those
related to the orbital periods. However, the limited time resolution
of our data does not allow us to rule out the occurrence of short period
oscillations with a time scale of the order of 10s like those
observed in DQ~Her type CVs or in some dwarf novae.

The light curve of variable V34=CV2 does not show any periodicity
although changes of $V$ magnitude exceeding 1 magnitude were observed on
several nights. The variable exhibited a likely dwarf nova type
outburst in the first half of  May 2003. At that time it  reached
$<V>\approx 16.5$,  which corresponds to $M_{V}\approx 4.3$ for an
assumed cluster membership. There is no evidence for the presence
of any superhumps during the event. Unfortunately  we have failed
to obtain any color information during the outburst.
%The observed amplitude of brightenning as well as absolute magnitude
%at maximum light are consistent with hypothesis that the variable
%exhibited a dwarf nova type outburst.

The variable V33=CV3 was observed at $19.1<V<19.9$ in 2002 and at
$18.3<V<19.5$ in 2004. On some nights non-periodic changes of $V$
exceeding 0.8 mag were seen. In May 2003 the star was found in a
low state with  $V>22.0$. Several  measurements for that run indicate
$V>22.5$ mag, although with  formal errors exceeding 0.4 mag.
Still, one may conclude that in May 2003 the variable was
certainly below $V=22.0$, corresponding to $M_{V}\geq 9.8$.

Grindlay et al. (1995) reported spectroscopic observations of
CV1-3 obtained with the $HST$ Faint Object Spectrograph.
They noted in particular the presence of He~II $\lambda 4686$ emission
lines in all three objects. That, in turn, lead to the conclusion that
CV1-3 are likely magnetic CVs. Our data are insufficient to rule out
the possibility that these systems indeed belong to the DQ Her type magnetic
systems. However, the presented light curves do not support the hypothesis
that any of the observed systems belongs to polars or intermediate polars.

\setcounter{figure}{3}
\begin{figure}
\begin{center}
\vspace{6.0cm}
\includegraphics{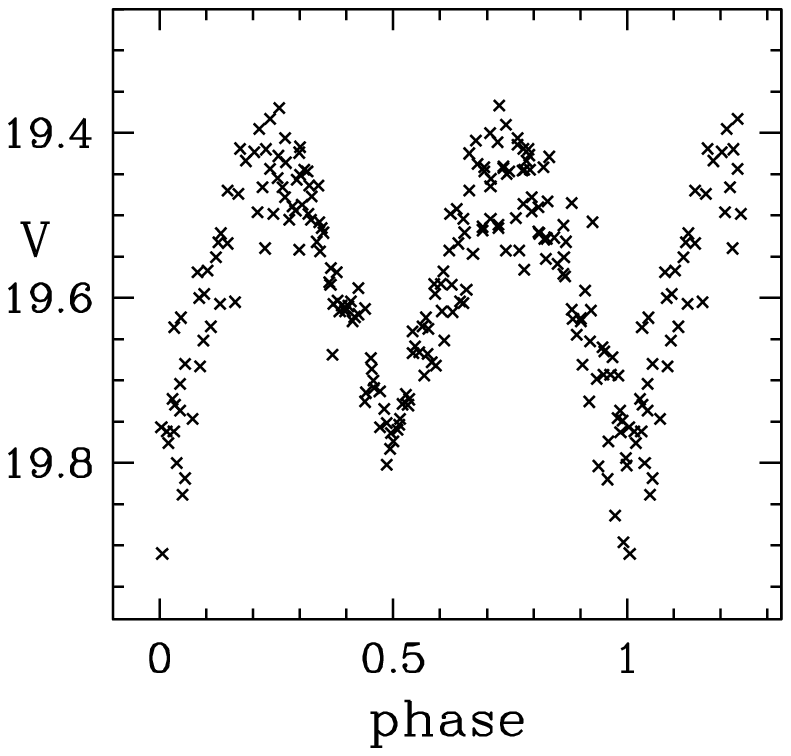}
\end{center}
\caption{\label{f4} Phased $V$ light curve of V13=CV6 for the
observing season 2004.}
\end{figure}

\setcounter{figure}{4}
\begin{figure}
\begin{center}
\vspace{9.4cm}
\includegraphics{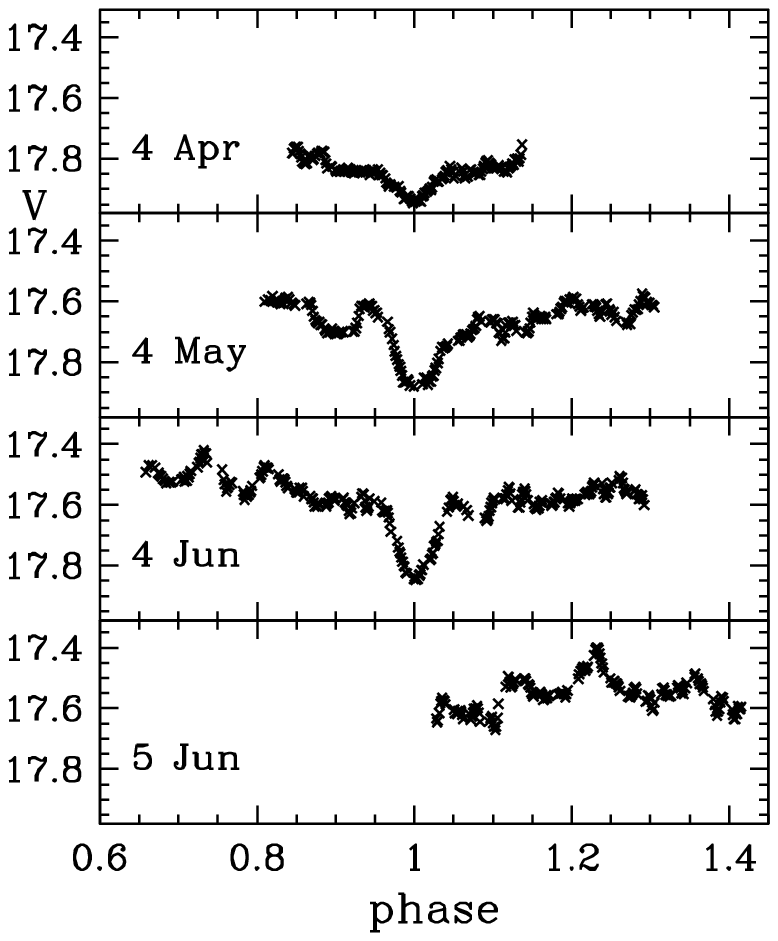}
\end{center}
\caption{\label{f5} Phased $V$ light curves of V12=CV1 for 4
nights from the observing season 2004.}
\end{figure}

%Continue to the end of page. Figures on next page.
%\clearpage

\setcounter{figure}{2}
\begin{figure*}
\begin{center}
\vspace{12.4cm}
%\special{psfile=fig3.ps hoffset=-55 voffset=-145 vscale=90 hscale=94}
\includegraphics{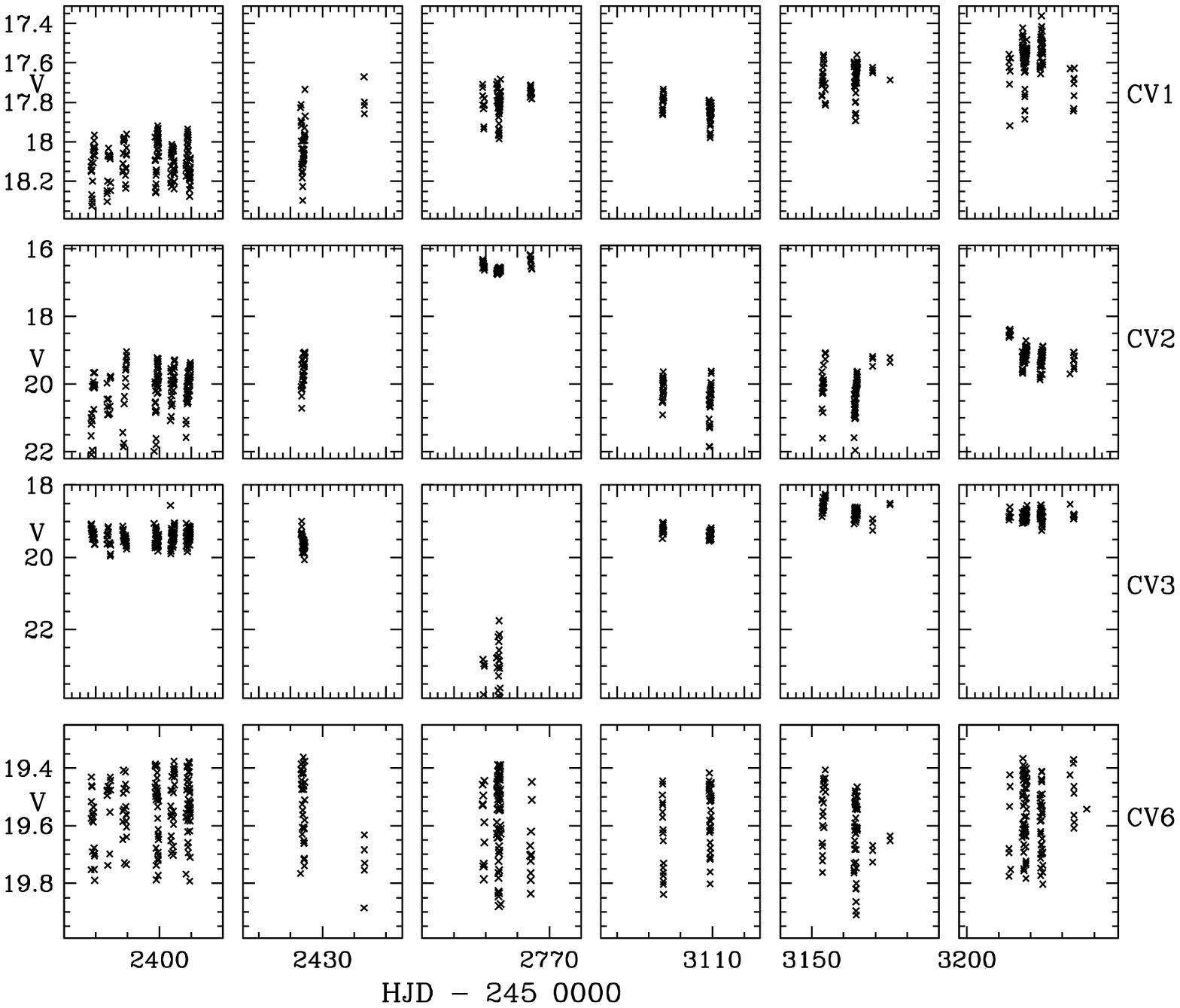}
\end{center}
\caption{\label{f3} Time domain $V$ light curves of observed CVs
for the period 2002-2004. Each box covers the time interval of 10
days. }
\end{figure*}

\subsection{Eclipsing binaries}
The main objective of our observing program was the detection of
detached eclipsing binaries belonging to the clusters. We hardly
succeeded in that respect despite quite extensive time coverage
(87 hours for the inner $8.65\times 2.6$~armin$^{2}$ field and 55
hours for the larger $8.65\times 4.3$~armin$^{2}$ field).  Two new
eclipsing  systems were identified.  A single eclipse event with
$\Delta V=0.5$~mag was observed for the variable V31. This object,
possessing an  X-ray counterpart, is most likely a foreground
binary of late spectral type. With $V_{max}=17.4$ and $B-V=1.28$ it
is located far to the red of the main sequence on the cluster
color-magnitude diagram. \\ A very likely eclipsing event with
$\Delta V\approx 0.03$~mag was detected for the variable V32. As
one can see in Fig. 6 a flat-bottomed part of eclipse and egress
were observed on the night of 11 April 2004. We checked carefully
that the observed event was not caused by any sort of instrumental
effects. There is also no evidence in our data that the photometry
of variable is affected by any unresolved blends reducing the observed
depth of a possible eclipse. Unfortunately, the variable is
located outside the fields covered by  $HST$ images currently
available publicly. Hence, we cannot exclude the possibility that, in
fact, the shallowness of the observed eclipse like event is due to
the presence of more than one star inside the fitted profile of V32.
The variable is located right at the top of the cluster turnoff on
the color-magnitude diagram (see Fig. 2). The light curve of V32,
including 32 hours of photometry from the 2002 season, can be
phased with a wide range of  periods but 10 shortest of them are
within a range 1.011038-1.060271~d. Other possible short periods
are grouped near 2.05 and 4.15~d. It is tempting to speculate that
the shallow transit like event observed in V32 is caused by a
large planet or brown dwarf orbiting the turnoff star from
NGC~6397. Such a possibility can be tested by spectroscopic
observations as well as by further photometry aimed at
determination of the orbital period and detailed analysis of shape
of the eclipse. In particular, determination of the orbital period
at $P\approx 1$~d would rule out a planetary transit hypothesis as
it implies too wide eclipses lasting about 0.1P. Spectroscopic
observations could on the other hand eliminate the possibility that
V32 is a hierarchical triple system including an eclipsing binary
(eg. Torres et al. 2004). \setcounter{figure}{5}
\begin{figure}
\begin{center}
\vspace{5.5cm}
%\special{psfile=fig6.ps hoffset=0 voffset=0 vscale=80 hscale=90}
\includegraphics{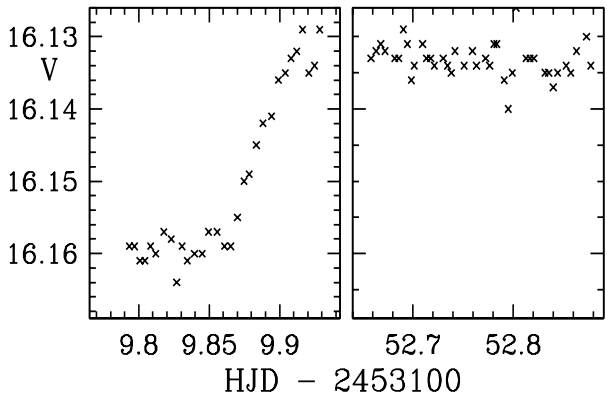}
\end{center}
\caption{\label{f6} $V$ band light curves of variable V32 obtained on nights
of 2004 April 11 and May 27.}
\end{figure}

\subsection{The optical companion to PSR J1740-5340}
The variable V16 is an optical companion to the millisecond pulsar
PSR J1740-5340 (D'Amico et al. 2001; Ferraro et al. 2001).
Its  variability is explained by the ellipsoidal effect.
The star has been recently a subject of detailed photometric and
spectroscopic studies (Kaluzny, Rucinski \& Thompson 2003;
Ferraro et al. 2003; Orosz \& van Kerkwijk 2003; Sabbi et al. 2003).
In particular, modeling of the light curve served as a basis for constraining
the orbital inclination of the system. It has been noted in
Kaluzny et al. (2003) that the optical  light curve of V16
exhibits some seasonal changes of its shape and amplitude. That claim is
further supported by our new data. The phased $V$ light and color
curves of V16 including the data from 2002-2004 seasons are
shown in Fig. 7.
Changes of the light curve level  reaching
0.03~mag  at a given phase are clearly visible. In particular we observed
the occurrence of changes with $\Delta V\approx 0.03$~mag within
the 2004 observing season spanning about 100 days.  The color curve
shows small orbital variations and the color becomes redder at
the upper and lower conjunctions (at phase 0.5 the observer sees the side
of the optical companion facing the pulsar). As  was already noted in
Ferraro et al. (2003), the lack of pronounced orbital  variations of the
color index is puzzling, considering detection  of high excitation
He~I absorption lines  in the spectrum of V16.
\setcounter{figure}{6}
\begin{figure}
\begin{center}
\vspace{7.6cm}
%\special{psfile=fig7.ps hoffset=-45 voffset=-367 vscale=100 hscale=90}
\includegraphics{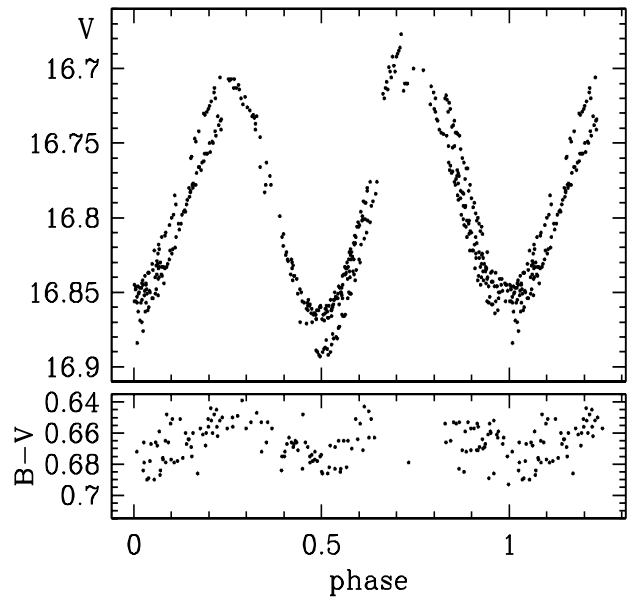}
\end{center}
\caption{\label{f7} Phased light and color curves of variable V16.}
\end{figure}

\subsection{SX Phe stars}
Detailed discussion of properties of SX~Phe variables from
NGC~6397 is beyond  the scope of this paper. We decided, however,
to provide a list of periods detectable in our time series
photometry. As it turns out, all the objects show multi-periodic
pulsations which  is a very common property of low amplitude SX~Phe
stars (eg. Rodr\'iguez \&  Breger 2001).\\ Analysis of observations
of SX Phe stars was repeated in two different ways: either by
iterations involving alternative period search and prewhitening of
the original data with all periods identified  so far, or by
successive prewhitening of one period at a time and period search
in the residuals. For period search we employed the multi-harmonic
analysis of variance periodogram (mhAoV, Schwarzenberg-Czerny,
1996). The periodogram is based on fitting data with an orthogonal
Fourier series. In the present case the series involved the
fundamental frequency and its first harmonic. This aided in
identification of the stronger, non-sinusoidal modes without
much affecting  detection of the faint, sinusoidal modes. The
advantage in employing  Fourier series, as opposed to a plain
sinusoid, is two-fold: harmonics tend to increase the detected
power and they no longer contribute to the residual noise. Fourier
series or phase binning in any periodogram produce sub-harmonics
as their side effect. These pose no practical difficulty as they
stand-out by their tight spacing of aliases.\\ Results of the
analysis are presented in Table 2 which for each star lists an
amplitude for the strongest mode of pulsation along with all
detected frequencies.
\begin{table*}
\centering
\begin{minipage}{200mm}
\caption{ List of frequencies identified in SX Phe stars.\label{t2}}
{\small
\begin{tabular}{llllllll}
\hline Star & $A_0$ & $f_0$ & $f_1$ & $f_2$ & $f_3$ & $f_4$ &
$f_5$ \\ \hline V10 & 0.0214(18) & 32.51591(8) & 33.30011(9) &
31.84787(12) &~&~&~\\ V11 & 0.0202(3) & 26.13539(2) & {\em
27.01135(8)$\dag$} & {\em 30.61296(14)$\dag$} &~&~&~\\ VV15 &
0.0135(5) & 46.61237(3)  & 42.05000(4) & {\em 39.79092(6)} & {\em
40.35747(7)} & 27.80867(6) & {\em 42.11866(10)}\\ V21 &
0.1351(17) & 25.66644(1) & 29.31764(10) &~&~&~&~\\ V23 &
0.0084(5) & 26.91564(5) & 26.26072(9) & 49.23396(12) & ~ & ~ &~\\
\hline
\end{tabular}}
\end{minipage}
{\small\noindent Remarks:
\begin{description}\item[{\em italics}] indicate
modes which yield clear spectral pattern, yet frequency is subject
to alias ambiguity;\item[$\dag$ cycle/year alias] ambiguity;
\end{description}}
\end{table*}
\subsection{Other variables}
The old variable V22 can be now classified as a double mode RRd
variable located in the cluster background. The available data allow
preliminary determination of the fundamental period and its first
overtone at $P0=0.52$~d and $P1=0.344$~d, respectively.\\ Four
objects from the sample of newly detected variables were
classified preliminarily as ellipsoidal variables. Their light
curves show periodic sinusoidal modulations. This is illustrated
in Fig. 8, which shows phased  light curves for the relevant stars. In
the case of V31 the observed scatter reflects mainly the change of 
average magnitude between 2003 and 2004 seasons. Spectroscopic data are
needed to verify the proposed classification and to establish
the membership status of these stars. \\ Figure 9 shows light curves of
V28 and V36 phased with the periods listed in Table 2. These two
stars are preliminarily classified as pulsating variables as
stability of their light curves seems to exclude belonging to spotted
variables.\\ The light curve of V25 for the 2004 observing season
showed periodic modulation with  $P=1.2306$~d. However, this
periodicity is absent in the data from the 2003 season.

\setcounter{figure}{7}
\begin{figure}
\begin{center}
\vspace{9.5cm}
\includegraphics{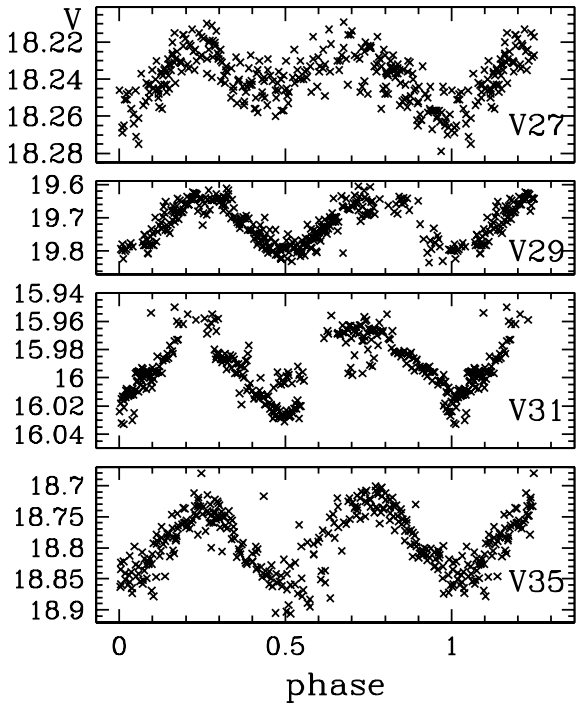}
\end{center}
\caption{\label{f7} Phased light curves of new candidate ellipsoidal variables.}
\end{figure}

\setcounter{figure}{8}
\begin{figure}
\begin{center}
\vspace{5.6cm}
\includegraphics{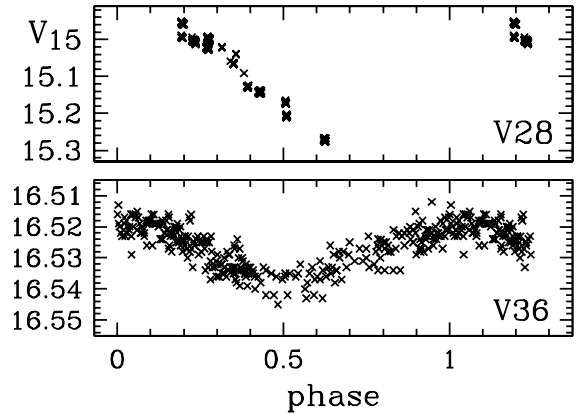}
\end{center}
\caption{\label{f7} Phased light curves of variables V28 and V36.}
\end{figure}

\section{Summary}\label{s6}
Our observing program was focused on the detection of  detached
eclipsing binaries in NGC~6397. Although two such objects were
identified, none of them can serve as an age or distance indicator
for the cluster. The binary V30 is most likely a foreground field
star. The second object, V32, has a good chance of being a cluster
member but  showed a single,  very shallow eclipse. It could be
composed of a cluster upper main sequence star orbited by a dim
companion being a large planet or brown dwarf. Another possibility
is that we have observed a grazing eclipse in an otherwise normal
binary. However, the shape of the light curve inside the observed eclipse
does not favor the latter alternative. At $V=16.1$ the star is
sufficiently bright for  spectroscopic follow-up
observations.\\ Our sample includes several candidates for
ellipsoidal variables. Some of them, primarily these located among
blue stragglers or close to the cluster main sequence, are also
interesting candidates for  spectroscopic follow-up. Because of
their location in post-core-collapse cluster some of them may be
detached binaries with degenerate components.\\ Based on 
observations from 3 observing seasons we have refined
the determination of orbital periods for two cataclysmic variables, CV1
and CV6. A very likely dwarf nova type outburst was detected for
the variable CV4. None of the four observed CVs showed any evidence
of  photometric variability pointing to its association with
magnetic systems. \\ Finally, our extensive and homogeneous
photometry shows unambiguously that the light curve of the optical
companion to the millisecond pulsar PSR J1740-5340 exhibits
seasonal changes of its amplitude.

\section*{Acknowledgments}
JK was supported by the grant 1~P03D 024 26 from the Ministry of
Scientific Research and Informational Technology, Poland.

\label{lastpage}


\begin{thebibliography}{99}
\bibitem[Alard \& Lupton, 1998]{ala98}Alard C., Lupton  R.~H.,
1998, ApJ, 503, 325
\bibitem[Alard, 2000]{ala00}Alard C., 2000, A\&AS, 144, 363.
\bibitem[Clement et al. 2001]{cle01}Clement C.~M. et al., 2001, AJ, 122, 2587
\bibitem[Cool et al. 1995]{coo95}Cool A., Grindlay J., Cohn H., Lugger P.,
Slavin S., 1995, ApJ, 439, 695
\bibitem[D'Amico et al. 2001]{dam01}D'Amico N., Lyne A.~G., Manchester
R.~N., Possenti A., Camilo F., 2001, ApJ, 548, L171
\bibitem[Ferraro et al. 2001]{fer01}Ferraro F.~R., Possenti A.,
D'Amico N., Sabbi E., 2001, ApJ, 561, L93
\bibitem[Ferraro et al. 2003]{fer03}Ferraro F.~R., Sabbi E., Gratton R.,
Possenti A., D'Amico N.~D., Bragaglia A., Camilo F.,
2003, ApJ, 584, L13
\bibitem[Grindlay et al., 1995]{gri95}Grindlay J.~E., Cool A.~M.,
Callanan P.~J., Bailyn C.~D., Cohn N.~H.,  Lugger P.M.,
1995, ApJ, 455, L47
\bibitem[Grindlay et al. 2001]{gri01}Grindlay J.~E., Heinke C.~O.,
Edmonds P.~D.,  Murray S.~S., Cool A.~M., 2001, ApJ, 563, 53
%\bibitem[Harris, 1996]{har96}Harris W.~E., 1996, AJ, 112, 1487
\bibitem[Kaluzny 1997]{kal97}Kaluzny J., 1997, A\&AS, 122, 1
\bibitem[Kaluzny \& Thompson, 2003]{kal03}Kaluzny J., Thompson I.~B.,
2003, AJ, 125, 2534 (KT03)
\bibitem[Kaluzny et al. 2003]{kal03a}Kaluzny J., Rucinski S.~M.,
Thompson I.~B., 2003, AJ, 125, 2534
\bibitem[Kaluzny et al. 2005]{kal05}Kaluzny, J., et al., 2005, in "Stellar
Astrophysics with the
World's Largest Telescopes", eds. J.~Mikolajewska \& A.~Olech, to
be published by AIP.
\bibitem[Mochejska et al. 2002]{moch02}Mochejska B.~J., Stanek K.~Z.,
Sasselov D.~D., Szentgyorgyi A.~H., 2002, AJ, 123, 3460
\bibitem[Orosz \& van Kerkwijk 2003]{orosz03}Orosz J.~A., van Kerkwijk
M.~H, 2003,  A\&A, 397, 237
\bibitem[Paczy\'nski 1997]{pac97}Paczy\'nski, B., 1997
Space Telescope Science Institute Series, The Extragalactic Distance Scale,
ed.  M. Livio (Cambridge University Press), 273
\bibitem[Piotto et al. 2002]{pio02}Piotto G. et al., 2002, A\&A, 391, 945
\bibitem[Sabbi et al. 2003]{sab03}Sabbi E., Gratton R., Ferraro F.~R.,
Bragaglia, A., Possenti A., D'Amico N., Camilo F., 2003, ApJ, 589, L41
\bibitem[Rodrígue \& Breger]{rod00}Rodr\'iguez E., Breger M., 2001, A\&A,
366, 178
\bibitem[Schwarzenberg, 1996]{alex96} Schwarzenberg-Czerny A., 1996, ApJ, 460, L107
\bibitem[Stetson, 1987]{ste87}Stetson P.~B., 1987, PASP, 99,  191
\bibitem[Stetson, 1990]{ste90}Stetson P.~B., 1990, PASP, 102, 932
\bibitem[Stetson, 2000]{ste00}Stetson P.~B., 2000, PASP, 112, 925
\bibitem[Torres, 2004]{tor04}Torres G., Konacki M., Sasselov D.~D.,
\& Jha, SJha S., 2004, ApJ, 614, 979
\end{thebibliography}
\end{document}